\documentclass[prl,twocolumn,showpacs]{revtex4}
\usepackage{amsfonts}
\usepackage{amsmath}
\begin{document}

\title{Efficient Classical Simulation of Optical Quantum Information
  Circuits}
\thanks{Published: Phys. Rev. Lett. \textbf{89}, 207903 (2002)}

\author{Stephen D. Bartlett}
\author{Barry C. Sanders} 
\affiliation{Department of Physics and Centre for Advanced Computing
  -- Algorithms and Cryptography,
  Macquarie University, Sydney, NSW 2109, Australia}

\begin{abstract}
  We identify a broad class of physical processes in an optical
  quantum circuit that can be efficiently simulated on a classical
  computer: this class includes unitary transformations,
  amplification, noise, and measurements.  This simulatability result
  places powerful constraints on the capability to realize exponential
  quantum speedups as well as on inducing an optical nonlinear
  transformation via linear optics, photodetection-based measurement
  and classical feedforward of measurement results, optimal cloning,
  and a wide range of other processes.
\end{abstract}
\date{October 30, 2002}
\maketitle

Quantum mechanics enables information processing that could not be
performed classically~\cite{Nie00}; examples include efficient
factorization and secure communication.  Optical realizations of
quantum information processes are particularly appealing because of
the robust nature of quantum states of light against the effects of
decoherence.  Both discrete-variable
(qubit)~\cite{Chu95,KLM01,Got01b} and continuous-variable
(CV)~\cite{Llo99} schemes offer significant potential for optical
quantum information processing, especially if efficient processes can
be performed that are not efficient on any classical device.

Advanced techniques in linear optics and squeezing, represented by
unitary transformations of optical states, are known to be
insufficient to implement arbitrary unitary
transformations~\cite{Llo99,Bar02b}.  In order to perform universal
quantum computation~\cite{Nie00}, optical nonlinear processes (such as
a Kerr nonlinearity~\cite{Wal94}) have been identified as a necessary
requirement both for qubit and CV schemes.  Nonlinear transformations
are also necessary for other optical quantum information processes
such as the Bell state measurements employed in quantum
teleportation~\cite{Lut99}.  Thus, the lack of a strong optical
nonlinearity with low losses greatly restricts the type of quantum
processes that can be performed in practice.  Recently, however,
nonunitary processes such as measurement have been identified as a
means to implement nonlinear operations.  Proposals for qubit-based
quantum computation~\cite{KLM01} and CV quantum
computation~\cite{Got01b} employ photon counting to induce nonlinear
transformations in optical systems.  Photon counting is an important
example of a process that can be used to achieve nonlinear
transformations via feedforward of measurement results.  Such a
nonunitary transformation appears to enable impressive capabilities
equivalent to nonlinear transformations.

In order to implement powerful optical quantum information processing
that cannot be performed classically, it is imperative to determine
what type of processes (unitary transformations, projective
measurements, interaction with a reservoir, etc.)\ can be used to
implement nonlinear transformations and thus perform universal quantum
computation.  One approach is to identify classes of processes that
can be efficiently simulated on a classical computer.  Under the
assumption that universal quantum computation is \emph{not}
efficiently simulatable classically, such processes are insufficient
to implement optical nonlinear transformations.  The Gottesman-Knill
(GK) theorem~\cite{Got99,Nie00} for qubits and the CV classical
simulatability theorem of Bartlett \emph{et al}.\ (BSBN)~\cite{Bar02b}
provide valuable tools for assessing the classical complexity
of a restricted class of controlled unitary transformations.

To develop a theorem that includes measurement and feedforward of
measurement results, positive operator-valued measures (POVMs) are
employed.  In related work, Knill~\cite{Kni01} has incorporated
certain projective measurements into a classical simulatibility result
for fermionic systems by allowing non-Hermitian Hamiltonians.  We
place this result in a generalized and powerful setting: unitary
transformations, POVMs and any other physical process can be described
in the unified formalism of completely positive (CP) maps.  We extend
the definition of the Clifford group of unitary transformations used
in the GK and BSBN theorems to the \emph{Clifford semigroup}; this
semigroup, expressed in the language of Gaussian CP maps, includes a
general class of unitary transformations, noise processes, amplifiers,
and measurements with feedforward.  Our result is a theorem for
efficient classical simulation of a broad class of optical quantum
information processes (both qubit and CV), and thus a no-go theorem
for universal optical quantum computation or inducing nonlinear
transformations.

Consider an optical quantum information process involving $n$ coupled
electromagnetic field modes, with a single mode described as a quantum
harmonic oscillator.  The two observables for the (complex) amplitudes
of the field mode serve as canonical operators for this oscillator.
For a system of $n$ coupled oscillators, the $2n$ canonical operators
$\{ \hat{q}_i,\hat{p}_i,i=1,\ldots,n\}$ satisfy
$[\hat{q}_i,\hat{p}_j] = i\hbar \delta_{ij} \hat{I}$, with $\hat{I}$
the identity.  We express the $2n$ canonical operators in the form of
phase space coordinates, defining $\hat{z}_i = \hat{q}_i$ and
$\hat{z}_{n+i} = \hat{p}_i$ for $i=1,\ldots,n$.  These operators
satisfy $[\hat{z}_i,\hat{z}_j] = i\hbar\Sigma_{ij}$, with $\Sigma$ the
skew-symmetric $2n \times 2n$ matrix $\Sigma_{ij} = \delta_{i+n,j} -
\delta_{i,j+n}$.  For a state $\rho$, the \emph{means} of the
canonical operators are defined to be $\xi_i = {\rm Tr}(\rho
\hat{z}_i)$, and the \emph{covariance matrix} is
\begin{equation}
  \label{eq:CovarianceMatrix}
  \gamma_{ij} = {\rm Tr}\bigl(\rho(\hat{z}_i - \xi_i)(\hat{z}_j -
  \xi_j) \bigr) -
  i \Sigma_{ij} \, .
\end{equation}
A \emph{Gaussian state} (a state whose Wigner function is Gaussian and
thus possesses a quasiclassical description) is completely
characterized by its means and covariance matrix~\cite{Lin00}.
Coherent states, squeezed states, and position- and
momentum-eigenstates are all examples of Gaussian states.

The Clifford group $\mathcal{C}_n$~\cite{Bar02b} is defined to be the
group of linear transformations of the canonical operators
$\{\hat{z}_i \}$.  For a system of $n$ oscillators, it is the unitary
representation of the group ISp($2n,\mathbb{R}$)
consisting of phase-space translations plus one- and two-mode
squeezing~\cite{Wun02}.  A displacement $X(\alpha)$ with $\alpha$ a
real $2n$-vector acts on the canonical operators in the Heisenberg
picture as
\begin{equation}
  \label{eq:Displacement}
  X(\alpha) : \hat{z}_i \to \hat{z}_i' = \hat{z}_i + \alpha_i \hat{I}
  \, .
\end{equation}
A symplectic transformation $M(A)$ acts as
\begin{equation}
  \label{eq:SymplecticTrans}
  M(A) : \hat{z}_i \to \hat{z}_i' =\sum_{k=1}^{2n} \hat{z}_k a_{ki} \,,
\end{equation}
with $A = (a_{ij})$ a real matrix satisfying $A^\dag \Sigma A =
\Sigma$.  A general element $C \in \mathcal{C}_n$ can be expressed as
a product $C(\alpha,A) = X(\alpha)M(A)$.

The Clifford group consists of unitary transformations that map
Gaussian states to Gaussian states; however, unitary transformations
do not describe all physical processes.  We define the \emph{Clifford
  semigroup}, denoted $\mathcal{K}_n$, to be the set of Gaussian CP
maps~\cite{Lin00} on $n$ modes: a Gaussian CP map takes any Gaussian
state to a Gaussian state.  Because Gaussian CP maps are closed under
composition but are not necessarily invertible, they form a semigroup.
A general element $T \in \mathcal{K}_n$ is defined by its action on
the canonical operators as
\begin{equation}
  \label{eq:ActionOfSemigroupOnCanonical}
  T(\alpha,A,G): \hat{z}_i \to \hat{z}_i' = \sum_k \hat{z}_k a_{ki} +
  \alpha_i \hat{I} + \hat{\eta}_i \, ,
\end{equation}
where $\alpha$ is a real $2n$-vector, $G = (g_{ij})$ is a $2n\times
2n$ real symmetric matrix, and $A$ is real but no longer required to
be symplectic.  Eq.~(\ref{eq:ActionOfSemigroupOnCanonical}) includes
both transformations (\ref{eq:Displacement}) and
(\ref{eq:SymplecticTrans}) plus additive noise processes~\cite{Gar00}
described by quantum stochastic noise operators $\{\hat{\eta}_i\}$
with expectation values equal to zero and covariance matrix
\begin{equation}
  \label{eq:NoiseCovariance}
  {\rm Tr}(\rho_R \hat{\eta}_i \hat{\eta}_j) - i \Sigma_{ij}=
  g_{ij} - i \sum_{kl} a_{ki} a_{lj} \Sigma_{kl} \, .
\end{equation}
Here, $\rho_R$ is a Gaussian `reservoir' state which, in order to
define a CP map, must be chosen such that the noise operators satisfy
the canonical quantum uncertainty relations.  This condition is
satisfied if the noise operators define a positive definite density
matrix,
\begin{equation}
  \label{eq:CPcondition}
  G +i \Sigma - i A^\dag \Sigma A \geq 0 \, .
\end{equation}
The Clifford group transformations are recovered for $G=0$.  (For
further details of Gaussian CP maps, see~\cite{Lin00}.)

The action of the Clifford semigroup on the means and covariance
matrix is given by
\begin{align}
  \label{eq:ActionOfSemigroupOnMeansAndCM}
  T(\alpha,A,G): \xi_i &\to \xi_i' = \sum_k \xi_k a_{ki} + \alpha_i
  \nonumber \\ 
  \gamma_{ij} &\to \gamma_{ij}' = \sum_{kl} a_{ki} \gamma_{kl}
  a_{lj} + g_{ij} \, . 
\end{align}
Because the means and covariance matrix completely define a Gaussian
state, the resulting action of the Clifford semigroup on Gaussian
states can be easily calculated via this action.

The Clifford semigroup $\mathcal{K}_n$ represents a broad framework to
describe several important types of processes in a quantum optical
circuit, as follows.

\emph{Linear optics and squeezing:} If $G=0$, then
condition~(\ref{eq:CPcondition}) demands that $A$ must be symplectic
and the corresponding map $T(\alpha,A,0)$ is $C(\alpha,A) \in
\mathcal{C}_n$.  The map $T(\alpha,A,0)$ is thus a unitary, invertible
transformation in the Clifford group.

\emph{Noise, amplification, and cloning:} The Clifford semigroup
possesses an illustrative interpretation in terms of a linear coupling
of the system's degrees of freedom to a reservoir.  Introduction of
noise into the quantum circuit can be incorporated using Clifford
semigroup maps with nonzero noise term $G$; such noise contributions
can be viewed as a coupling of system modes to reservoir modes in
Gaussian states via a beamsplitter or mode coupler.  Linear
amplifiers, including phase-insensitive and phase-sensitive
amplifiers, are also describable in this context, with $A$
characterizing the amplifying term and $G$ the introduction of the
associated noise~\cite{Cav82}.  Linear amplifiers can be used to
perform optimal cloning of Gaussian states~\cite{Lin00,Bra01},
describable using Gaussian CP maps~\cite{Lin00}.

\emph{Measurements in the Clifford semigroup:} The Clifford semigroup
encompasses a broad class of measurement.  Measurements as CP maps are
most easily illustrated using the Kraus operator sum~\cite{Kra83}.  A
general CP map $\mathcal{E}$ is defined by its action on an $n$-mode
state $\rho$ as $\mathcal{E}(\rho) = \sum_k O_k \rho O_k^\dag$, where
$\{ O_k \}$ are operators on the $n$-mode Hilbert space.  For a
measurement process, the operators $\{ O_k \}$ are the elements of a
POVM, and the index $k$ labels a specific measurement outcome.  As an
example, let the operators $O_k$ be projections $|\alpha_k\rangle_i
\langle\alpha_k|$ onto a complete set $\{|\alpha_k\rangle_i\}$ of
coherent states for the $i$th mode.  The properties of coherent states
ensure that the resulting map is Gaussian CP.  In the continuous
limit where the sum $\sum_k$ is replaced by the integral $\int {\rm
  d}^2 \alpha/\pi$, it describes the projective measurements
corresponding to eight-port homodyne detection~\cite{Art65} on mode
$i$.  If the measurement result is discarded (i.e., the reservoir,
describing the measurement apparatus, is not observed), this process
is described by $\mathcal{E}$ mapping a pure state to a mixture of
possible measurement outcomes.

However, if the measurement outcome $\alpha_k$ is recorded, the
corresponding CP map describing the measurement is different, because
the density matrix is updated to reflect our informed knowledge of the
measurement outcome.  In our example, measuring a specific result
$\alpha_k$ leads to
\begin{equation}
  \label{eq:MeasuredOutcome}
  \rho \to \rho' = \frac{|\alpha_k\rangle_i \langle \alpha_k
  |\rho|\alpha_k \rangle_i \langle \alpha_k|}{{\rm Tr}\bigl(|\alpha_k
  \rangle_i \langle \alpha_k |\rho\bigr) } \, . 
\end{equation}
This map is also Gaussian CP.  The $i$th mode is left in the known
state $|\alpha_k\rangle_i$, and the remaining modes are collapsed into
the corresponding multimode state, renormalized by the probability of
the measurement outcome $\alpha_k$.

The projections need not be onto coherent states for the measurement
to be in the Clifford semigroup; projections onto squeezed states are
also Gaussian CP.  In the limit of infinite squeezing, it could also
be a projection onto quadrature-phase eigenstates via homodyne
detection~\cite{Yue80}.  Projective measurements onto multimode
entangled Gaussian states~\cite{Wan02} (e.g., Einstein-Podolsky-Rosen
(EPR) states) are also Gaussian CP.  Furthermore, measurements need
not be projective; Gaussian POVMs can be constructed by adding ancilla
modes according to Neumark's theorem~\cite{Per93}.  Finally, noisy
measurement can be described using the Clifford semigroup by
composition with a linear noise map as described previously.

\emph{Conditional transformations:} Clifford semigroup maps
conditioned on classical numbers or the outcome of such measurements
can also be described using Gaussian CP maps.  Because of the
composition property of Clifford semigroup maps, any Gaussian CP map
conditioned on the outcome of a Gaussian CP measurement will also be
Gaussian CP.

We now present the primary result of this paper, which is the
capability to simulate processes in the Clifford semigroup efficiently
on a classical machine.  Recall that any Gaussian state is completely
characterized by its means and covariance matrix.  For any quantum
information process that initiates in a Gaussian state and involves
only Clifford semigroup maps, one can follow the evolution of the
means and the covariance matrix rather than the quantum state itself.
For a system of $n$ coupled oscillators, there are $2n$ independent
means and $2n^2+n$ elements in the (symmetric) covariance matrix;
thus, following the evolution of these values requires resources that
are polynomial in the number of coupled systems.

\textbf{Theorem:} \textit{Any quantum information process that
  initiates in a Gaussian state and that performs only Clifford
  semigroup maps can be \emph{efficiently} simulated using a classical
  computer.}  These maps include (i) unitary Clifford group
transformations (displacements and squeezing), (ii) linear
amplification (including phase-insensitive and phase-sensitive
amplification and optimal cloning), linear loss mechanisms or additive
noise, (iii) Clifford semigroup measurements including, but not
limited to, projective measurements in the position/momentum
eigenstate basis or coherent/squeezed state basis, with finite
losses, and (iv) Clifford semigroup maps conditioned on classical
numbers or the outcomes of Clifford semigroup measurements (classical
feedforward).

This theorem extends beyond the simulatability results of GK and BSBN.
Both of these previous theorems follow the evolution of the Pauli
operators under unitary Clifford group transformations and projective
measurements in the computational basis, allowing for the efficient
simulation of processes initiated in the computational basis.  By
instead following the means and covariance matrix of Gaussian states,
this new theorem allows for the simulation of a much broader class of
initial states, nonunitary transformations and measurements to be
included beyond that of GK and BSBN while still incorporating all of
their results.

We now consider some of the key new results of this theorem in terms
of known processes.  One is that optimal cloning of Gaussian states
cannot be used to advantage for any exponential quantum computational
speedup.  Also, any Clifford semigroup transformation conditioned on
the measurement outcome of homodyne detection with finite losses using
Gaussian states cannot be used to induce a nonlinear transformation,
nor can a projection onto a multimode Gaussian state be used for this
purpose.  In terms of optical implementations of quantum computing,
this theorem reveals why all previous schemes either propose some form
of optical nonlinearity~\cite{Chu95,KLM01,Got01b,Llo99} or are not
efficiently scalable~\cite{Cer98}.

This theorem places severe constraints on the use of photodetection to
perform nonlinear transformations in both qubit~\cite{KLM01} and
CV~\cite{Got01b} realizations of optical quantum computing.  For a
threshold photodetector~\cite{Bar02a} with perfect efficiency, the
POVM is given by two elements, corresponding to ``absorption'' and
``non-absorption'' of light; these elements are
\begin{equation}
  \label{eq:PhotodetectionPOVM}
  \Pi_{0} = |0\rangle \langle 0| \, , \quad
  \Pi_{>0} = \sum_{n=1}^{\infty} |n\rangle\langle n| \, ,
\end{equation}
where $|n\rangle$ are Fock states of definite photon number $n$.
Photon counters are effectively constructed as arrays of such
detectors~\cite{Bar02a}.  The vacuum projection describes the
non-absorption measurement, and the corresponding map describing this
measurement result is Gaussian CP.  However, the absorption outcome is
not.  As a result, $\mathcal{K}_n$ elements conditioned on the
no-absorption outcome of a photodetection measurement are in
$\mathcal{K}_n$, whereas transformations conditioned on the absorption
outcome are not.  Note that the same result holds for
finite-efficiency photodetectors: such detectors can be modelled as
perfect efficiency photodetectors with a linear loss
mechanism~\cite{Yue80} describable (see above) using the Clifford
semigroup.  Thus, the absorption outcome of photodetection and the
feedforward of this measurement result is a key resource for optical
quantum information processing.

As an example, consider the generation of single photon Fock states
via parametric down conversion (PDC)~\cite{Har96}.  The transformation
corresponding to PDC is two-mode squeezing; thus, the production of
the squeezed vacuum is in the Clifford semigroup.  The use of a
photodetector in one mode (described by
Eq.~(\ref{eq:PhotodetectionPOVM})) can be used to post-selectively
create non-Gaussian states (approximately single photon states) in the
other mode conditioned on the absorption measurement outcome; such a
process, then, is not Gaussian CP.  The no-absorption measurement is
in the Clifford semigroup, which leaves the other mode in a Gaussian
state (the vacuum).  Thus, the creation or use of single-photon Fock
states lies outside the domain of this theorem.

Schemes for employing photon counting for linear optical quantum
computation are thus constrained by two results of this theorem.
First, linear optics gates conditioned on the non-absorption
measurement of a vacuum cannot be used to induce a nonlinear
transformation.  The other constraint is that any nonlinear gate
employing linear optics and photon counting \emph{must} be
nondeterministic; a photon counting measurement of a Gaussian state
could possibly result in an outcome of zero photons, and such a result
corresponds to an efficiently classically simulatable process.  (Note
that nonlinear optics, in contrast, can be deterministic.)

The quantum search algorithm of Grover~\cite{Gro97} does not yield an
exponential speedup over classical search algorithms.  It is
nevertheless interesting to investigate if this algorithm can be
efficiently simulated using the methods presented here.  We note that
linear optical implementations~\cite{Kwi00} of Grover's algorithm
satisfy the conditions of our theorem and are thus simulatable;
however, the resource requirements of these implementations are not
scalable~\cite{Blu02}.  It is not known if a scalable optical
realization would require extra resources (i.e., an optical
nonlinearity).

Our theorem for efficient classical simulation provides a powerful
tool in assessing whether a given optical process (such as photon
counting) can enhance linear optics to perform nonlinear
transformations or allow quantum processes that are exponentially
faster than classical ones.  Algorithms or circuits employing a large
class of CP maps given by the Clifford semigroup can be efficiently
simulated on a classical computer, and thus do not provide any sort of
quantum exponential speedup~\cite{Eis02}.  Many quantum optics
experiments can be described in terms of the Clifford semigroup; thus,
the challenge is to develop and exploit techniques that lie outside
the Clifford semigroup and may be used to realize powerful quantum
information processes in an optical system.

\begin{acknowledgments}
  We acknowledge helpful discussions with S.\ L.\ Braunstein, J.\ 
  Eisert, E.\ Knill, and M.\ B.\ Plenio.  This project has been
  supported by an ARC Large Grant and a Macquarie University New Staff
  Grant.
\end{acknowledgments}

\end{document}